\documentclass[conference,letterpaper]{IEEEtran}

\usepackage{fancyhdr} 
\usepackage{amssymb,amsmath} 
\usepackage{graphicx}   
\usepackage{subfigure}  
\usepackage[boxed,figure,noline]{algorithm2e}   

\begin{document}

\title{Multi-Channel Deficit Round-Robin Scheduling for Hybrid TDM/WDM Optical
  Networks}

\author{
  \IEEEauthorblockN{Mithileysh Sathiyanarayanan and Kyeong Soo Kim}
  \IEEEauthorblockA{College of Engineering, Swansea University\\
    Swansea, SA2 8PP, United Kingdom\\
    Email:\{m.sathiyanarayanan.611702, k.s.kim\}@swansea.ac.uk}
}

\maketitle


\begin{abstract}
  In this paper we propose and investigate the performance of a multi-channel
  scheduling algorithm based on the well-known deficit round-robin (DRR), which
  we call multi-channel DRR (MCDRR). We extend the original DRR to the case of
  multiple channels with tunable transmitters and fixed receivers to provide
  efficient fair queueing in hybrid time division multiplexing (TDM)/wavelength
  division multiplexing (WDM) optical networks. We take into account the
  availability of channels and tunable transmitters in extending the DRR and
  allow the overlap of `rounds' in scheduling to efficiently utilize channels
  and tunable transmitters. Simulation results show that the proposed MCDRR can
  provide nearly perfect fairness with ill-behaved flows for different sets of
  conditions for interframe times and frame sizes in hybrid TDM/WDM optical
  networks with tunable transmitters and fixed receivers.
\end{abstract}



\section{Introduction}
Scheduling is a method of harmonizing the access to system resources among
competing data flows. It is achieved by specifying the order and the allotted
time period for packets from each flow. Scheduling is an important part of
networking systems because it not only enables the sharing of the bandwidth but
also guarantees the quality of service (QoS). Well-designed scheduling
algorithms could provide higher throughput, lower latency, and better fairness
with lower complexity in serving the packets. As such, the scheduling plays an
important role in achieving high performance of the networking systems.

Due to its importance in networking and communication, the scheduling has been
extensively studied but mainly in the context of single-channel
communication. The advent of wavelength division multiplexing (WDM) technology,
however, demands the extension of this packet scheduling problem to the case of
multi-channel communication, especially with tunable transmitters for hybrid
time division multiplexing (TDM)/ wavelength division multiplexing (WDM) systems. 
The main objective of the multi-channel scheduling is to schedule the transmissions 
of the data over multiple channels to the users. The important measures in choosing a scheduling
algorithm are throughput, latency, fairness, and complexity. The major focus of
existing work is mostly on the throughput and delay performance of the
scheduling algorithm like in SUCCESS-HPON \cite{Kim:05-1}, but there is hardly
any support for fairness guarantee. The main objective of this paper is to study
the multi-channel scheduling in hybrid TDM/WDM optical networks with tunable
transmitters and fixed receivers providing fairness in throughput. In this paper
we propose and investigate the performance of a multi-channel deficit
round-robin (MCDRR) scheduling algorithm which can provide throughput fairness
among flows with different size packets with $O(1)$ processing per packet.

As for multi-channel scheduling with tunable transceivers in hybrid TDM/WDM
optical networks, the work for the SUCCESS-HPON architecture in \cite{Kim:05-1}
provides a detailed investigation of several multi-channel scheduling algorithms
like batching earliest departure first (BEDF) and sequential scheduling with
schedule time framing (S$^3$F) under realistic environments, which is one of the
basis for the work in this paper. Through extensive simulations using tunable
transmitters and receivers, it has been demonstrated that both the BEDF and
S$^3$F improves the throughput and delay performances. Note that we consider the
case of tunable transmitters and fixed receivers in this paper, while the
SUCCESS-HPON architecture is based on both tunable transmitters and tunable
receivers.

The proposed MCDRR is based on the deficit round-robin (DRR) scheduling
algorithm which extends the simple round-robin with deficit counters
\cite{Shreedhar:95-1}. The DRR provides good fairness, lower complexity, and
lower implementation cost, which makes it an ideal candidate for high-speed
gateways or routers.

In the basic DRR scheme, stochastic fair queuing (SFQ) \cite{mckenney91:_stoch}
is used to assign flows to queues. For serving the queues, round-robin
scheduling is used, with a quantum of service assigned to each flow. The DRR
scheduler in rotation selects packets to send out from all flows that have
queued. The DRR maintains a service list to keep the flow sequence being served
in a round and to avoid examining empty queues. It differs from the traditional
round-robin in that if a queue is unable to send a packet in the previous round
because a packet was too large, the remainder from the previous quantum is added
to the quantum for the next round. Queues that are not completely serviced in a
round are compensated in the next round.

During each round, a flow can transmit at once as many packets as possible if
there is enough quantum for them. For each flow, two variables --- i.e., quantum
and deficit counter --- are maintained. Quantum is the amount of credits in
bytes allocated to a flow within the period of one round. Deciding the quantum
size is an important issue. If we expect that the work for DRR is $O(1)$ per
packet, then the quantum for a flow should be larger than the maximum packet
size from the flow so that at least one packet per backlogged flow can be served
in a round \cite{Shreedhar:95-1}.

Note that the multi-channel scheduling have been studied by others in slightly
different contexts that ours: For instance, optimal wavelength scheduling for
Hybrid WDM/TDM Passive Optical Networks (PONs)
\cite{wang11:_optim_wavel_sched_hybrid_wdm} inspects the upstream wavelength
scheduling in hybrid wavelength division multiplexing and time division
multiplexing passive optical networks (WDM/TDM PONs), where the minimum resource
allocation unit is a time slot on a wavelength. They use three optimal
wavelength scheduling algorithms for the three kinds of hybrid WDM/TDM PONs.
\begin{itemize}
\item Type-I WDM/TDM PONs: Each optical network unit (ONU) has a single optical
  transmitter with a tunable wavelength.
\item Type-II WDM/TDM PONs: Each ONU still has a single transmitter, but some
  are fixed to transmit at a certain wavelength.
\item Type-III WDM/TDM PONs: Each ONU has one or more transmitters and all
  transmitters can tune their wavelengths.
\end{itemize}
They proposed algorithms based on the round-robin scheduling (RRS) and shortest
channel first (SCF) concept to calculate the optimal schedule length and achieve
the best wavelength scheduling with the shortest schedule length and the maximum
channel utilization.
Also, to provide fairness guarantee with multiple channels, the extension of
fair queueing (FQ) has been studied in \cite{Blanquer:01-1,Cobb:02-1}, but they
are for fixed transceivers as in static WDM systems. The closest to our work in
this paper is the study of multi-server round-robin scheduling in
\cite{xiao04:_analy_of_multi_server_round}. Unlike the hybrid TDM/WDM optical
network where a specific wavelength is dedicated to a specific destination,
however, they assume that flows can use any of multiple channels.

Our paper is mainly based on the tunable transmitters and fixed receivers in the
multi-channel system which requires investigation in the performance of a
multi-channel deficit round-robin (MCDRR) scheduling algorithm, which can
provide fairness (in terms of throughput) for flows with different size packets
with O(1) processing per packet.

The rest of the paper is organized as follows. In Section II we explain the
concept of packet service in rounds in the multi-channel case and explain the
enqueueing and dequeuing processes in detail in the MCDRR algorithm. We also
illustrate the MCDRR algorithm with examples. In Section III we present
simulation results for the MCDRR algorithm. Section IV concludes our discussions
in this paper.

\section{Multi-Channel Deficit Round-Robin (MCDRR)}
\begin{figure}[hbtp]
  \centering
  \includegraphics*[angle=-90,width=\linewidth]{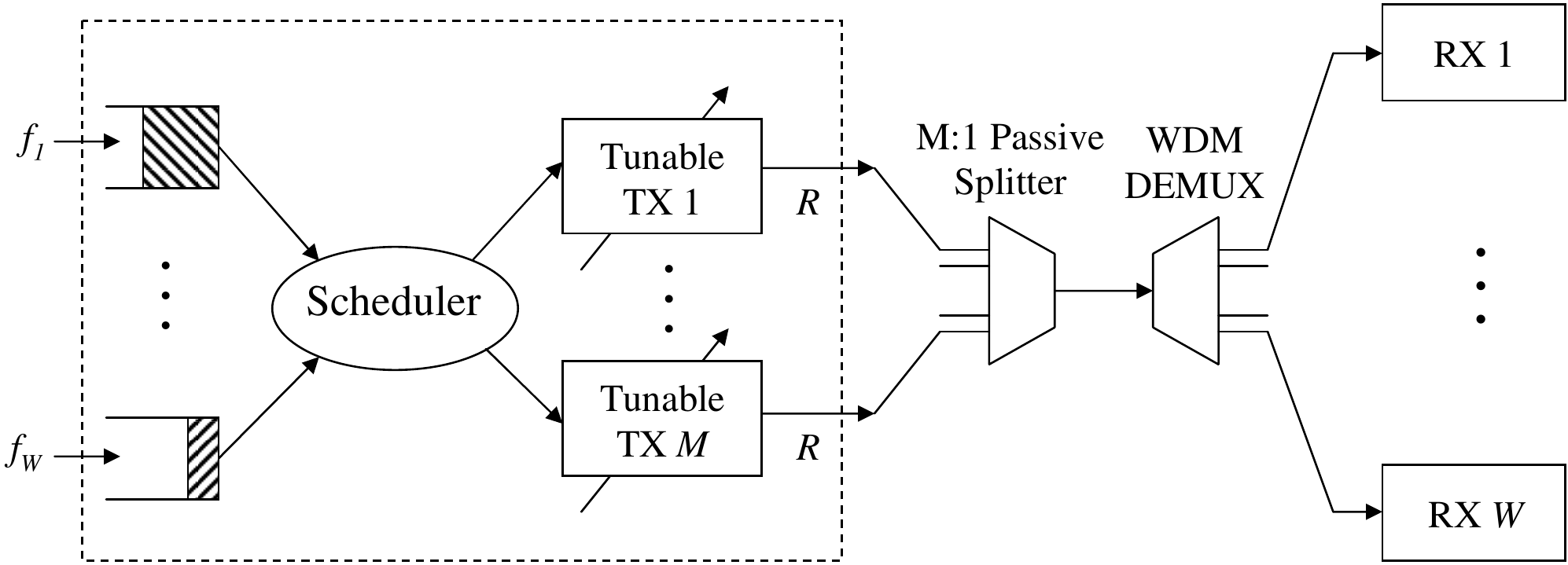}  
  \caption{Block diagram of a hybrid TDM/WDM link based on tunable transmitters
    and fixed receivers.}
  \label{fig:hybrid_link}
\end{figure}

The scheduling of packets in switches and routers has been studied mainly in the
context of single-channel communication with fixed transceivers. We extend the
packet scheduling problem to the case of multi-channel communication with
tunable transmitters and fixed receivers as shown in
Fig. \ref{fig:hybrid_link}.\footnote{That is also a model for the downstream
  links of future hybrid TDM/WDM PON with tunable transmitters at the optical
  line terminal (OLT) and fixed receivers at the optical network units (ONUs).}
The proposed MCDRR, the multi-channel extension of the DRR, takes into account
the availability of channels and tunable transmitters and overlaps `rounds' in
scheduling to efficiently utilize channels and tunable transmitters. To service
the queues (i.e., virtual output queues (VOQs)), we use the simple round-robin
algorithm with a quantum of service assigned to each queue as in the case of
DRR.

Because the MCDRR allows multiple rounds to overlap and run in parallel, the
scheduling and the transmission of packets are not necessarily sequential unlike
the DRR. To take into account these parallel operations and their timing
relations, therefore, we need a precise description of the MCDRR scheduling
algorithm and provide a detailed pseudocode in Fig. \ref{fig:mcdrr_pseudocode}.
\begin{algorithm}[hbtp]
  \SetKw{Initialization}{Initialization}
  \Initialization{}\;
  \For{$i \leftarrow 0$ \KwTo $W-1$}{
    $DC[i]=0$\;
  }
  \BlankLine
  \SetKw{Arrival}{Arrival}
  \Arrival{on the arrival of a packet $p$ from channel $i$}\;
  \If{Enqueue(i, p) is successful}{
    \If{a transmitter is available}{
      $(ptr,~ch) \leftarrow Dequeue()$\;
      \If{pkt $\neq$ NULL}{
        $Send(*ptr,~ch)$\;
        \lIf{VOQ[i] is empty}{$DQ[i] \leftarrow 0$\;}
      }
    }
  }
  \BlankLine
  \SetKw{Dequeue}{Dequeue}
  \Dequeue{}\;
  $startQueueIndex \leftarrow (currentQueueIndex + 1) \% W$\;
  \For{$i \leftarrow 0$ \KwTo $W-1$}{
    $idx \leftarrow i + startQueueIndex \% W$\;
    \If{$VOQ[idx]$ is not empty}{
      $DC[idx] \leftarrow DC[idx] + Q[idx]$\;
      \If{$numPktsScheduled[idx] == 0$}{
        $currentQueueIndex \leftarrow idx$\;
        $pos \leftarrow 0$\;
        $ptr \leftarrow \&packet(VOQ[idx],~pos)$\;
        \Repeat{$DQ[idx] \geq length(*ptr)$}{
          $DC[idx] \leftarrow DC[idx] - length(*ptr)$\;
          $numPktsScheduled[idx]++$\;
          $pos++$\;
          $ptr \leftarrow \&packet(VOQ[idx],~pos)$\;
          \lIf{ptr is NULL}{Exit the loop\;}
        }
        $Return~(\&packet(VOQ[idx],~0),$
        $~~~~~~currentQueueIndex)$\;
      }
    }
  }
  $Return~NULL$\;
  \BlankLine
  \SetKw{Departure}{Departure}
  \Departure{at the end of transmission on channel $i$}\;
  $numPktsScheduled[i]--$\;
  \If{$numPktsScheduled[i] >0$}{
    $ptr \leftarrow \&packet(VOQ[i],~0)$\;
    $Send(*ptr,~i)$\;
    \lIf{VOQ[i] is empty}{$DC[i] \leftarrow 0$\;}
  }
  \Else{
    $(ptr,~ch) \leftarrow Dequeue()$\;
    \If{$ptr \neq NULL$}{
      $Send(*ptr,~ch)$\;
      \lIf{VOQ[ch] is empty}{$DC[ch] \leftarrow 0$\;}
    }
  }
  \caption{Pseudocode for the MCDRR algorithm.}
  \label{fig:mcdrr_pseudocode}
\end{algorithm}

$Enqueue(i,~p)$ is a standard queue operation to put a packet $p$ into a VOQ for
channel $i$. $Dequeue()$ is a key operation of the MCDRR scheduling and returns
a pointer to the head-of-line (HOL) packet in the selected VOQ or $NULL$ when
the scheduler cannot find a proper packet to transmit. $packet(queue,~pos)$
returns a pointer to the packet at the position of $pos$ in the $queue$ or
$NULL$ when there is no such packet.

For each $VOQ[i]$, we maintain the following counters:
\begin{itemize}
\item $DC[i]$: It contains the byte that $VOQ[i]$ did not use in the previous
  round.
\item $numPktsScheduled[i]$: It counts the number of packets scheduled for
  transmission during the service of $VOQ[i]$. Unlike the original DRR, we need
  this counter to keep track of those packets scheduled for transmission due to
  multiple rounds overlapped and running in parallel.
\end{itemize}

\subsection{MCDRR Example}

\begin{figure}[hbtp]
\centering
\resizebox{0.50\textwidth}{!}{\includegraphics[trim=20mm 0mm 20mm 0mm,clip=true,angle=-90,scale=1]{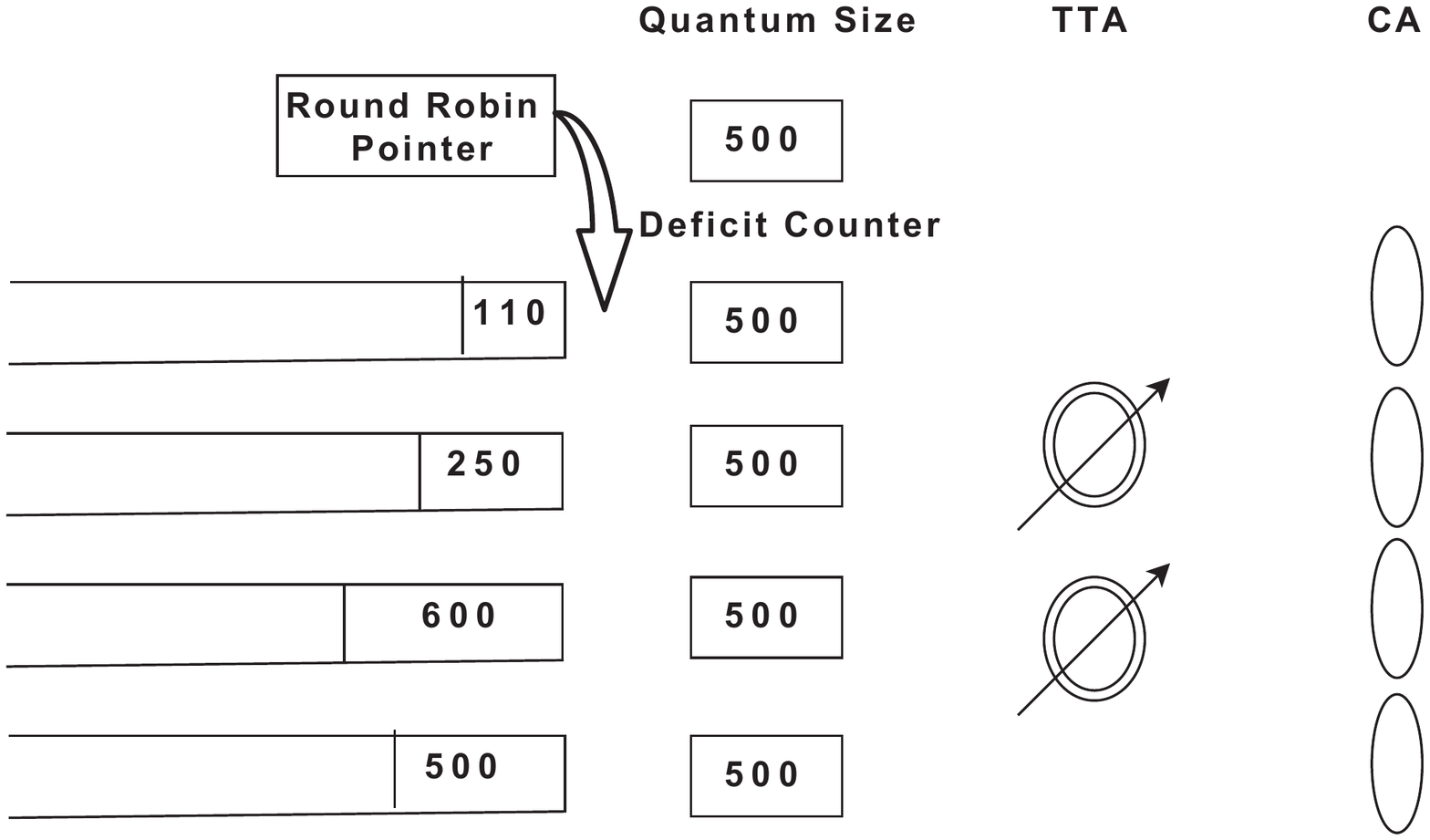}}
\caption{Start of Round 1}
\end{figure}


\begin{figure}[hbtp]
\centering
\resizebox{0.50\textwidth}{!}{\includegraphics[trim=20mm 0mm 20mm 0mm,clip=true,angle=-90,scale=1]{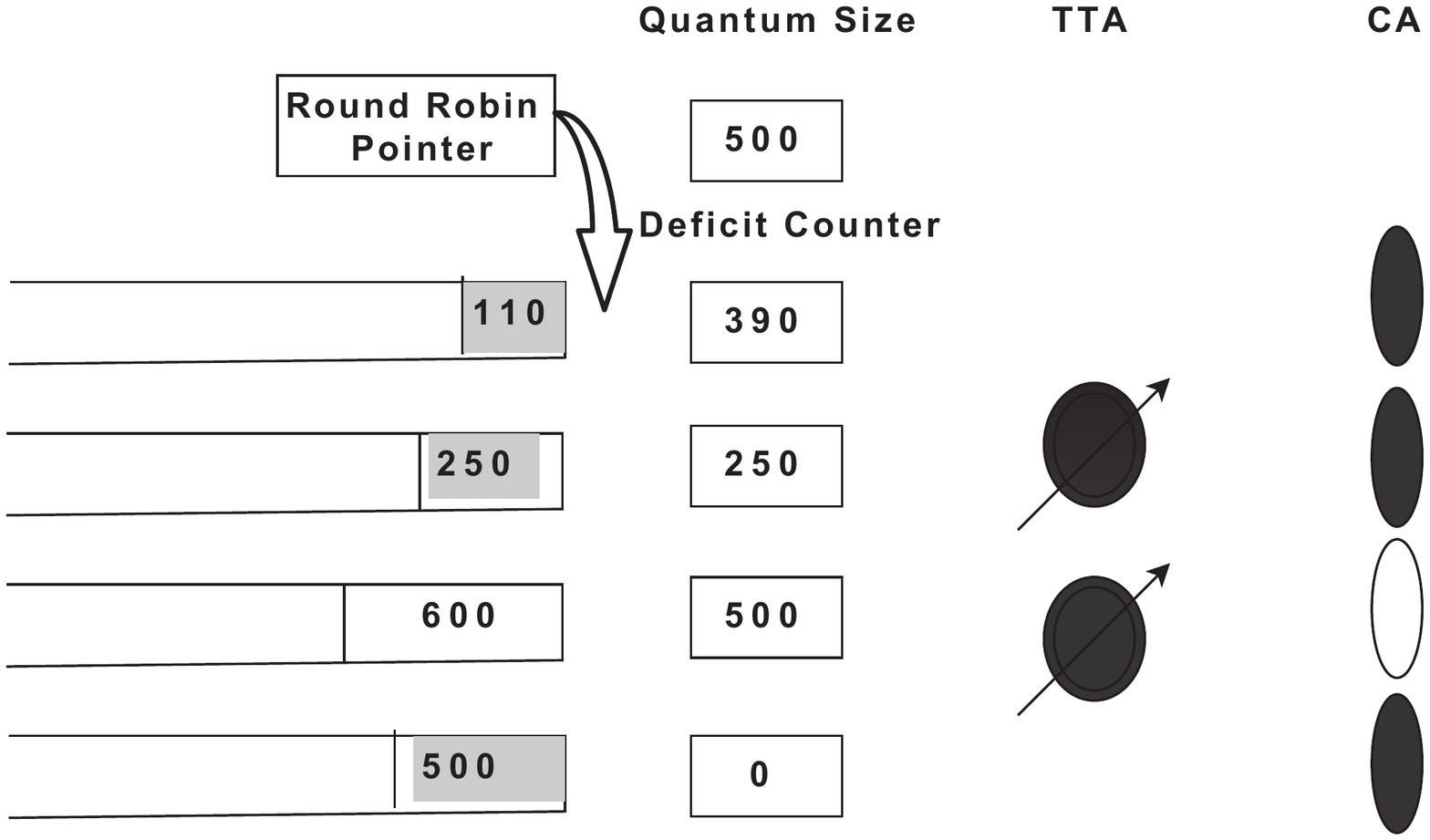}}
\caption{End of Round 1}
\end{figure}

\begin{figure}[hbtp]
\centering
\resizebox{0.50\textwidth}{!}{\includegraphics[trim=20mm 0mm 20mm 0mm,clip=true,angle=-90,scale=1]{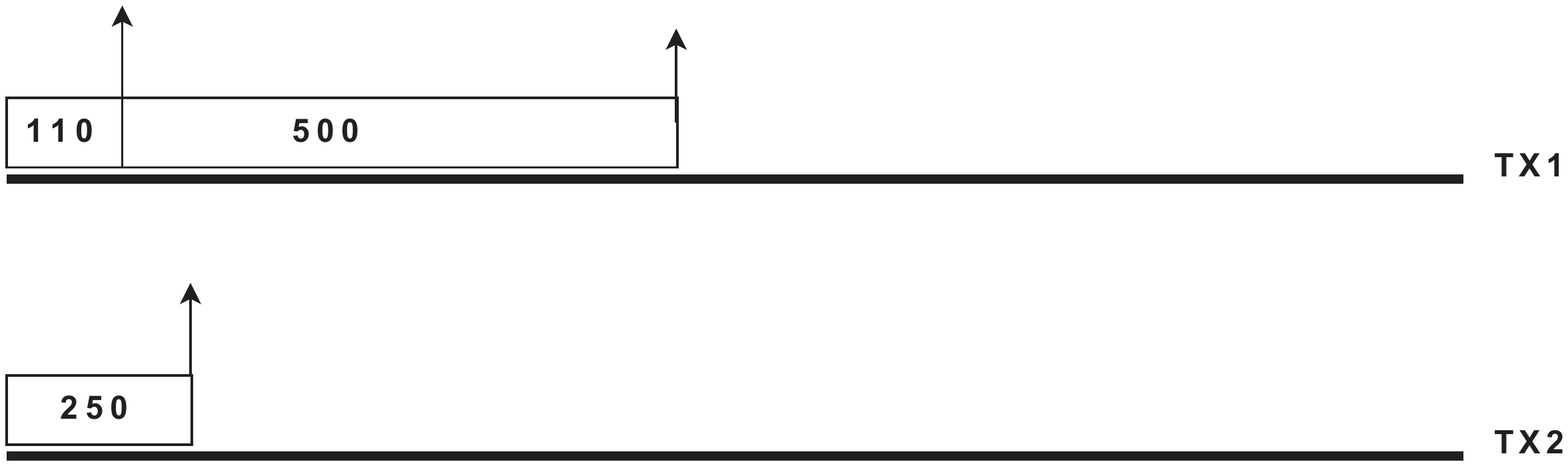}}
\caption{End of Round1 showing in Tunable Transmitter}
\end{figure}

\begin{figure}[hbtp]
\centering
\resizebox{0.50\textwidth}{!}{\includegraphics[trim=20mm 0mm 20mm 0mm,clip=true,angle=-90,scale=1]{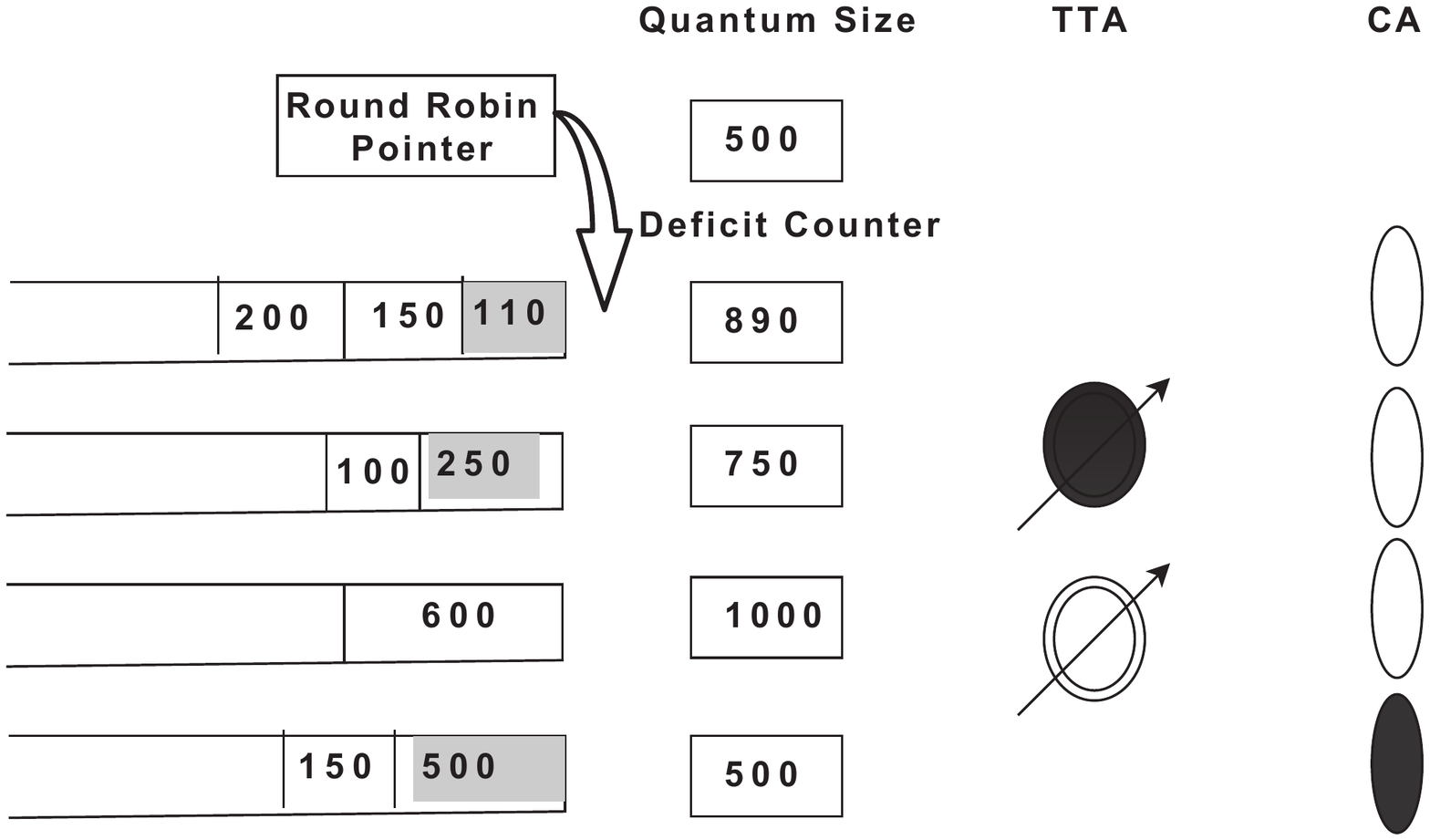}}
\caption{Start of Round 2}
\end{figure}

\begin{figure}[hbtp]
\centering
\resizebox{0.50\textwidth}{!}{\includegraphics[trim=20mm 0mm 20mm 0mm,clip=true,angle=-90,scale=1]{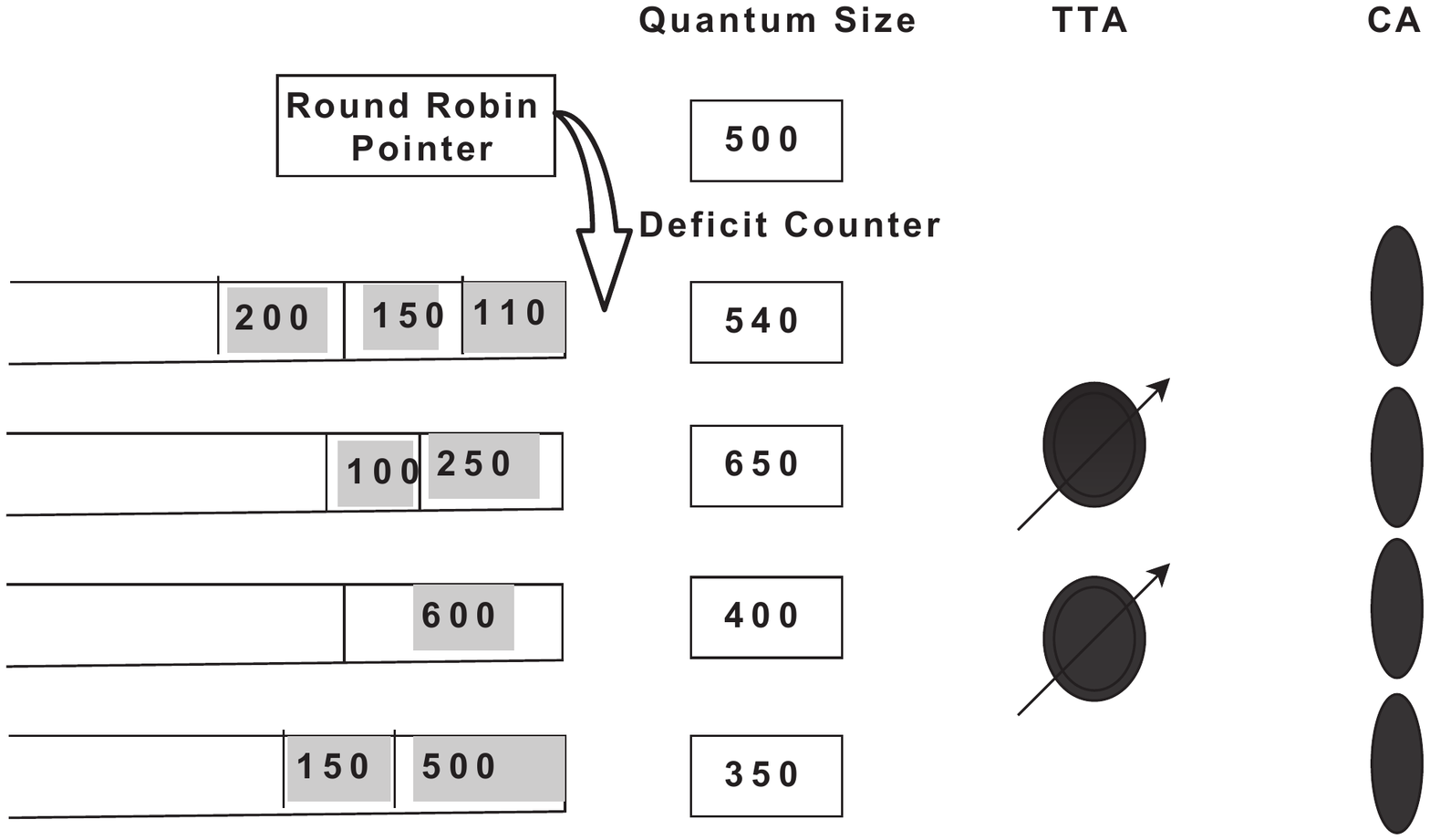}}
\caption{End of Round 2}
\end{figure}

\begin{figure}[hbtp]
\centering
\resizebox{0.50\textwidth}{!}{\includegraphics[trim=20mm 0mm 20mm 0mm,clip=true,angle=-90,scale=1]{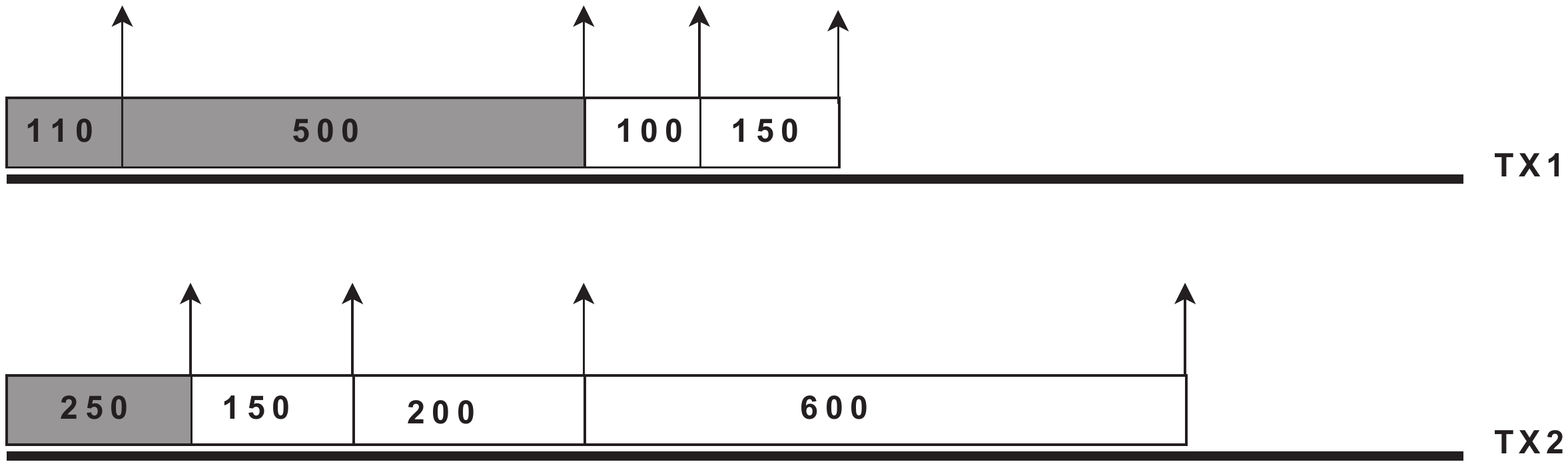}}
\caption{End of Round2 showing in Tunable Transmitter}
\end{figure}

\begin{figure}[hbtp]
\centering
\resizebox{0.50\textwidth}{!}{\includegraphics[trim=20mm 0mm 20mm 0mm,clip=true,angle=-90,scale=1]{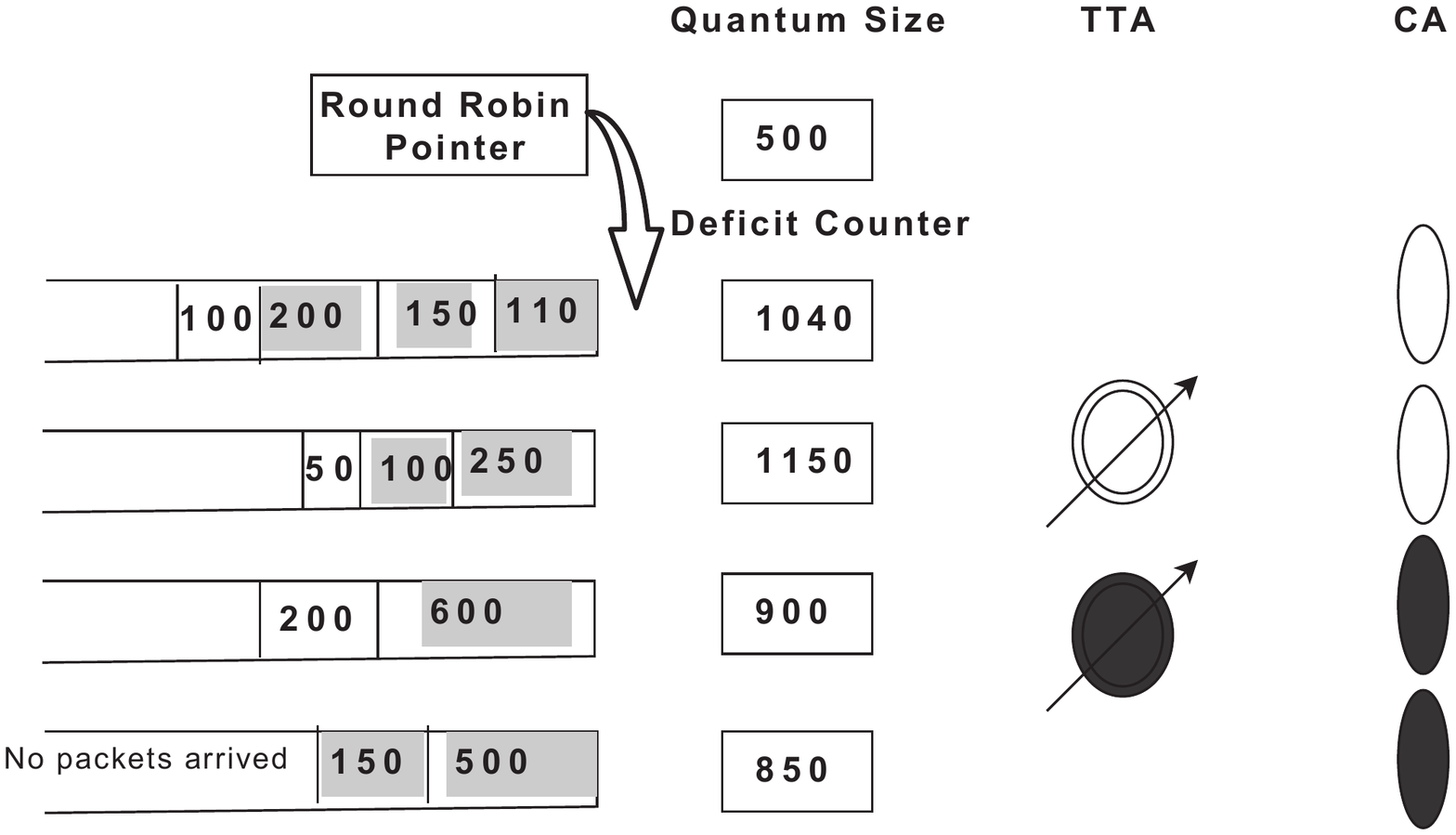}}
\caption{Start of Round 3}
\end{figure}

\begin{figure}[hbtp]
\centering
\resizebox{0.50\textwidth}{!}{\includegraphics[trim=20mm 0mm 20mm 0mm,clip=true,angle=-90,scale=1]{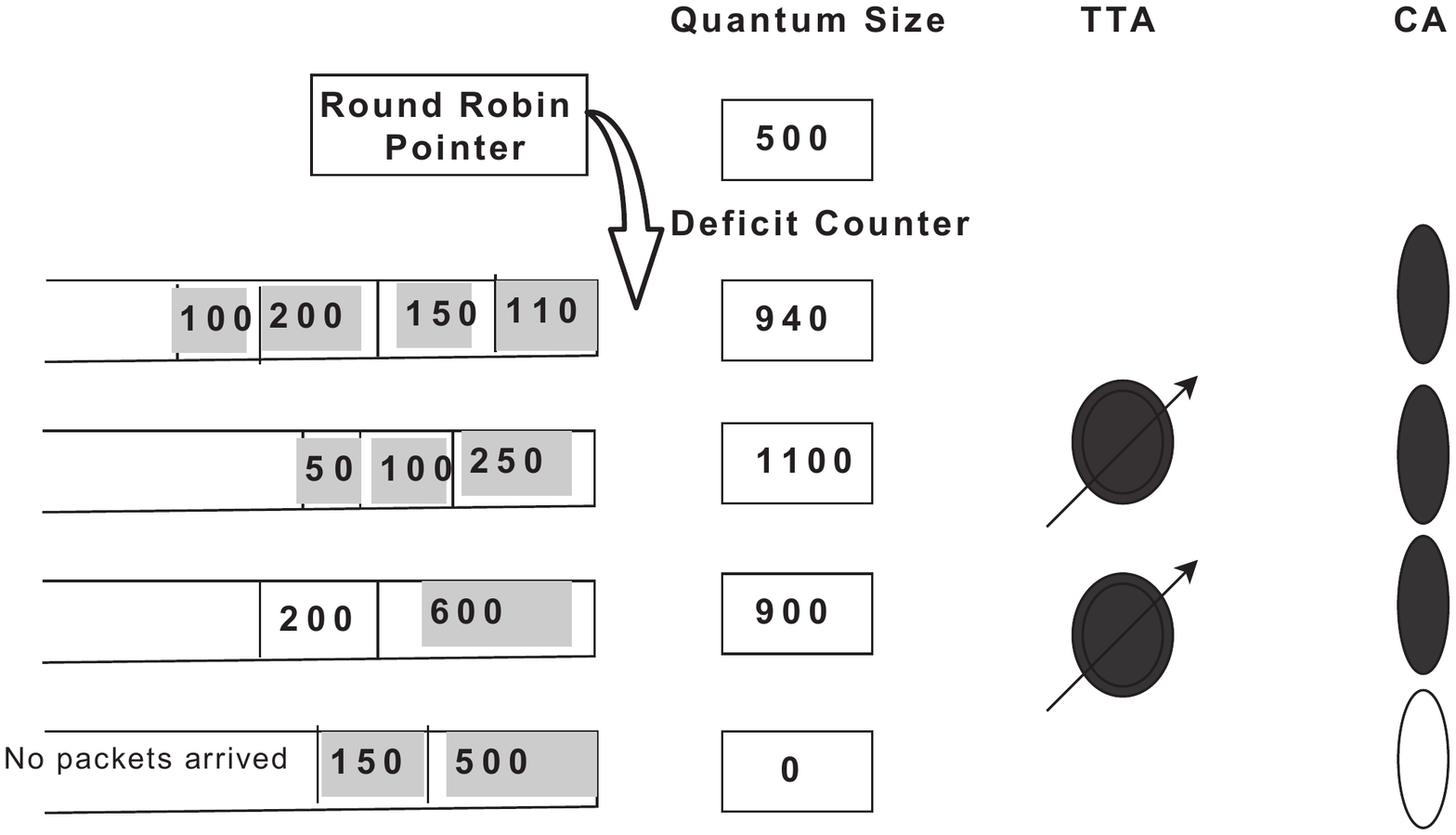}}
\caption{End of Round 3}
\end{figure}

\begin{figure}[hbtp]
\centering
\resizebox{0.50\textwidth}{!}{\includegraphics[trim=20mm 0mm 20mm 0mm,clip=true,angle=-90,scale=1]{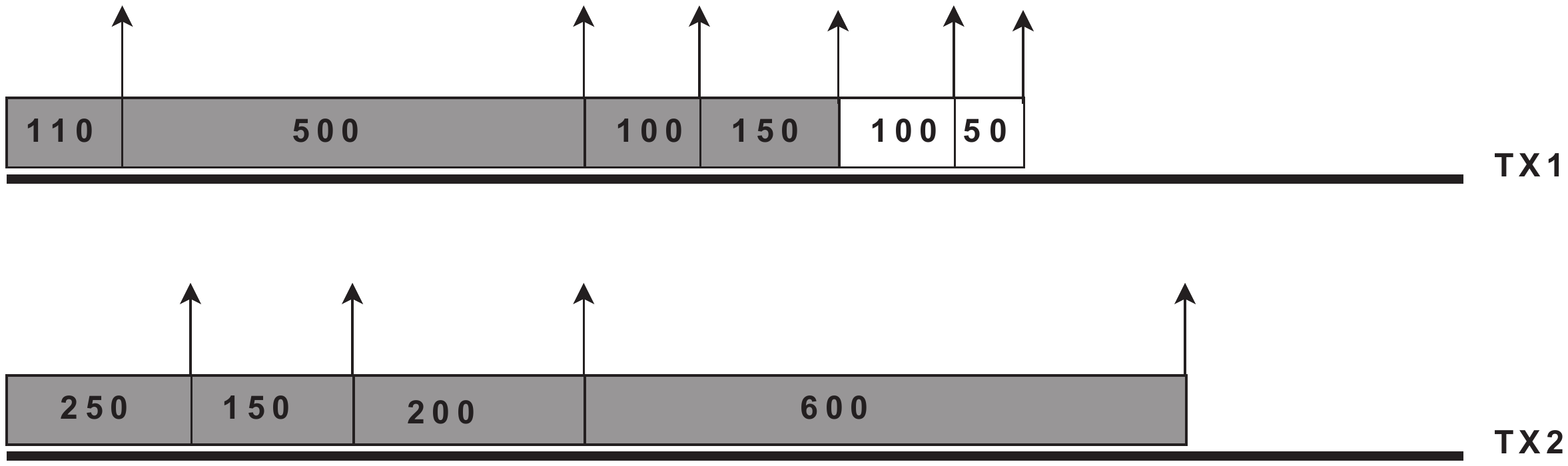}}
\caption{End of Round3 showing in Tunable Transmitter}
\end{figure}

\begin{figure}[hbtp]
\centering
\resizebox{0.50\textwidth}{!}{\includegraphics[trim=20mm 0mm 20mm 0mm,clip=true,angle=-90,scale=1]{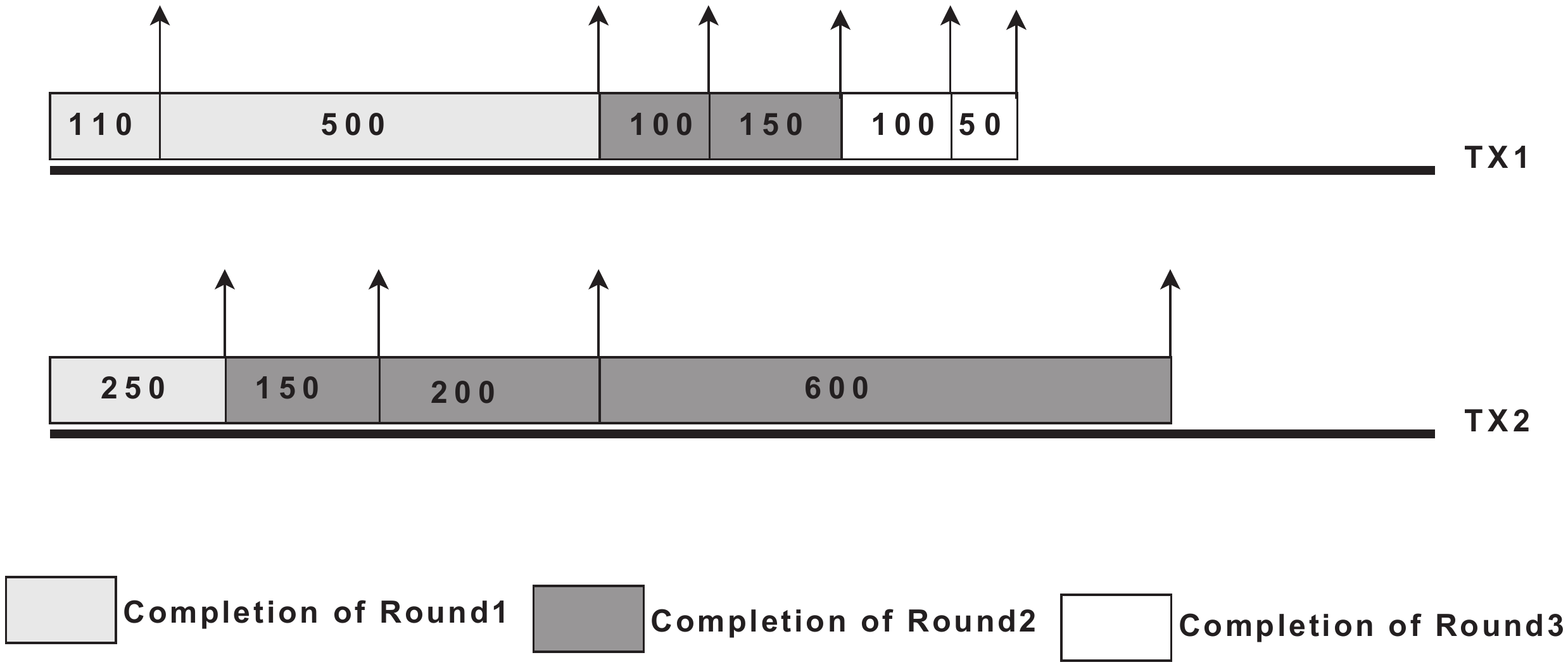}}
\caption{Overlapping of all the three rounds in MCDRR Scheduling}
\label{fig:overlapping}
\end{figure}

The arrows in the above diagrams shows the triggering of scheduling process
after the transmission of packet from each flow. After serving packets from each
flow, the tunable transmitter triggers the scheduling process. The
Fig. \ref{fig:overlapping} shows the overlapping of rounds. The MCDRR is carried
out in such way, where the next round starts as the previous round still in
progress. It means that the delay is avoided and the channel does not remain
idle when packets satisfy all the criteria.

\subsection{MCDRR Example Explanation}
At the start of the First Round, the tunable transmitter available triggers the
scheduling process. The round robin pointer starts from the first flow
initialized. The deficit counter becomes equal to the quantum size. If the
packet size is lesser than the deficit counter and channel is available at that
instant of time, the packet is served. If the channel is not available, the
pointer is moved to the next flow. When the channel becomes available then the
packet will be transmitted in the next round.

In the example quantum size is considered to be 500 credits, now both the
tunable transmitters are available, the pointer starts from the Flow 1, the
packet of size 110 bytes will be served since it is less than the deficit
counter 500 credits and the channel is available at that instant of time. By
default they choose tunable transmitter 1. After serving, the deficit counter is
updated, that is DC becomes 390 credits.

Since the tunable transmitter 2 is also available, the pointer moves to the Flow
2, the packet of size 250 bytes will be served since they are less than the
deficit counter 500 credits and the channel is available at that instant of
time. DC is updated. Now TX1 becomes available and triggers the scheduling
process, the pointer moves to Flow 3, the packet size is greater than the
deficit counter and the flow is skipped. DC remains the same. Still the TX1 is
available, so the packet in Flow 4 of size 500 bytes is served
successfully. Since the pointer has moved through all the given flows, we say it
as ``Completion of one Round''.

Now TX2 becomes available and triggers the scheduling process, which is the
start of next Round, that is Second Round. The deficit counter is updated with
the quantum size again i.e quantum is added to all the deficit counters of the
respective flows. DC= DC(prev) + Quantum Size. The pointer starts from the Flow
1 again. The DC becomes 390 credits + 500 credits. In this second round, two
packets in Flow 1 had arrived. According to our description, only one packet can
be served from each flow irrespective of packet size as far as they satisfy the
dequeuing criteria. So the packet size of 150 bytes can be served successfully
with channel available at that instant but not the packet of 200 bytes (because
only one packet can be served per flow as per our description). After the
service, the DC is updated again.

At some instant, both the tunable transmitters TX1 and TX2 can be available. In
that case by default TX1 will be chosen. In this case TX2 triggers again, the
packet of 100 bytes in Flow 2 is served successfully. The packet in Flow 3 which
was not served in previous round is been served in this round using TX2 because
the DC is 1000 credits now. Now TX1 becomes available and the packet of 150
bytes in Flow 4 is served, that is the End of Round.

The TX1 becomes available and triggers the scheduling process, which is the
start of the next round, that is the Third Round. The DC is updated with the
quantum size again. The pointer starts from the Flow 1. The packet of 200 bytes
in the Flow 1 is served and after some instant again TX1 becomes available and
the packet in Flow 2 is served. Since the packet size is small, TX1 becomes
available at the earliest compared to the TX2. The packet (200 bytes) in Flow 3
cannot be served at this instant though the tunable transmitter is available
because channel is not available, it means that the packet is still being served
from the previous round. So the pointer moves to the next flow and packets
arrived in this round means no packets to be served in this Third round i.e Flow
is empty, in that case DC is reset to zero for fairness issues. So that is the
end of this round. Since TX1 is still available, the next round starts from the
Flow 1 and the process continues till all the flows completely become empty
which is sequential.

This example covers all the details such as packet size lesser than the deficit
counter with channel available and channel not available at some instant of
time, then packet size greater than the deficit counter with channel available
and not available, flow being empty in one particular round.





\section{Simulation Results}

\begin{figure*}[hbtp]
  \begin{minipage}{.48\linewidth}
    \begin{center}
     \includegraphics*[width=\linewidth]{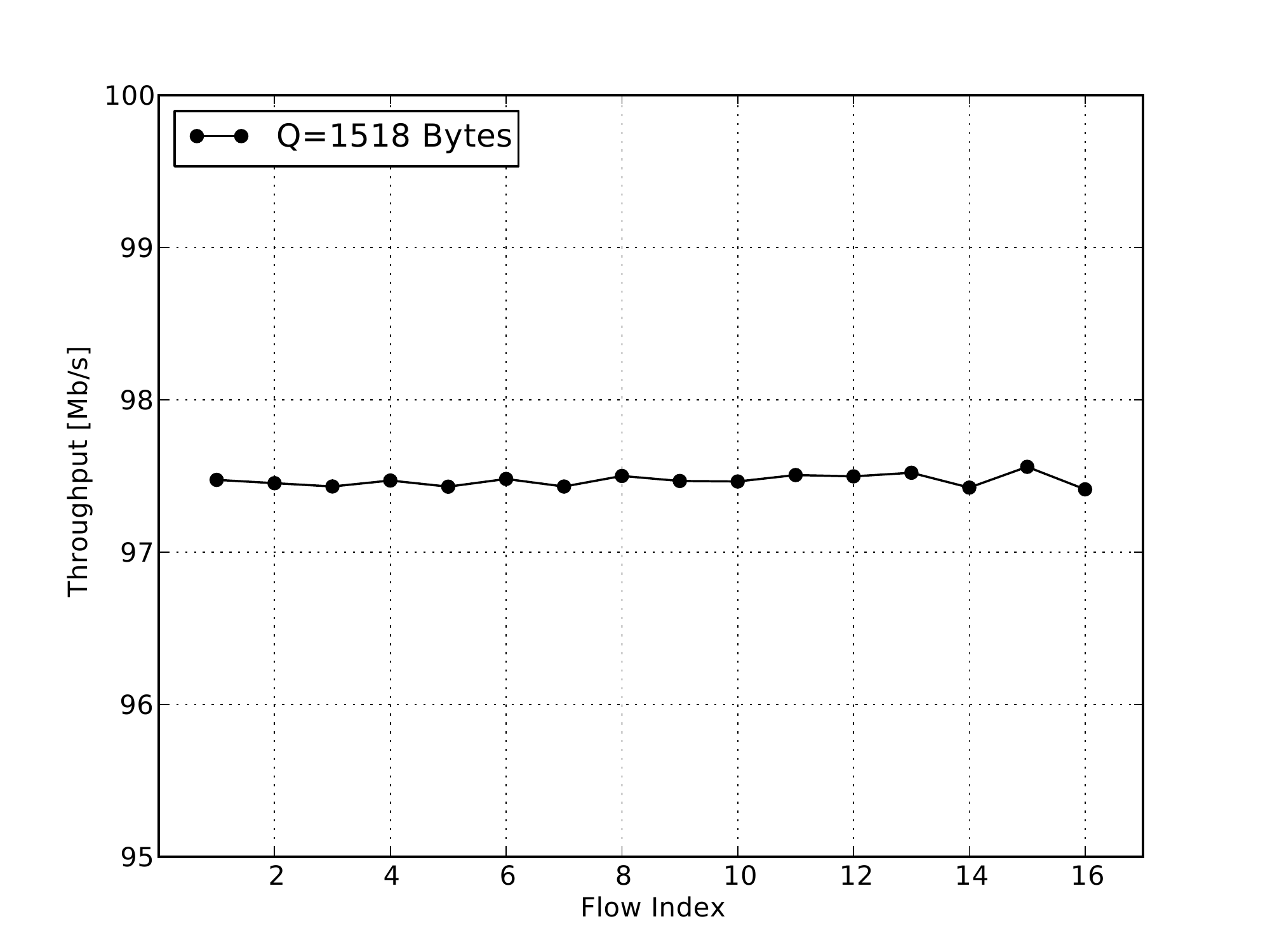}\\
      {\scriptsize (a)}
    \end{center}
  \end{minipage}
  \hfill
  \begin{minipage}{.48\linewidth}
    \begin{center}
      \includegraphics*[width=\linewidth]{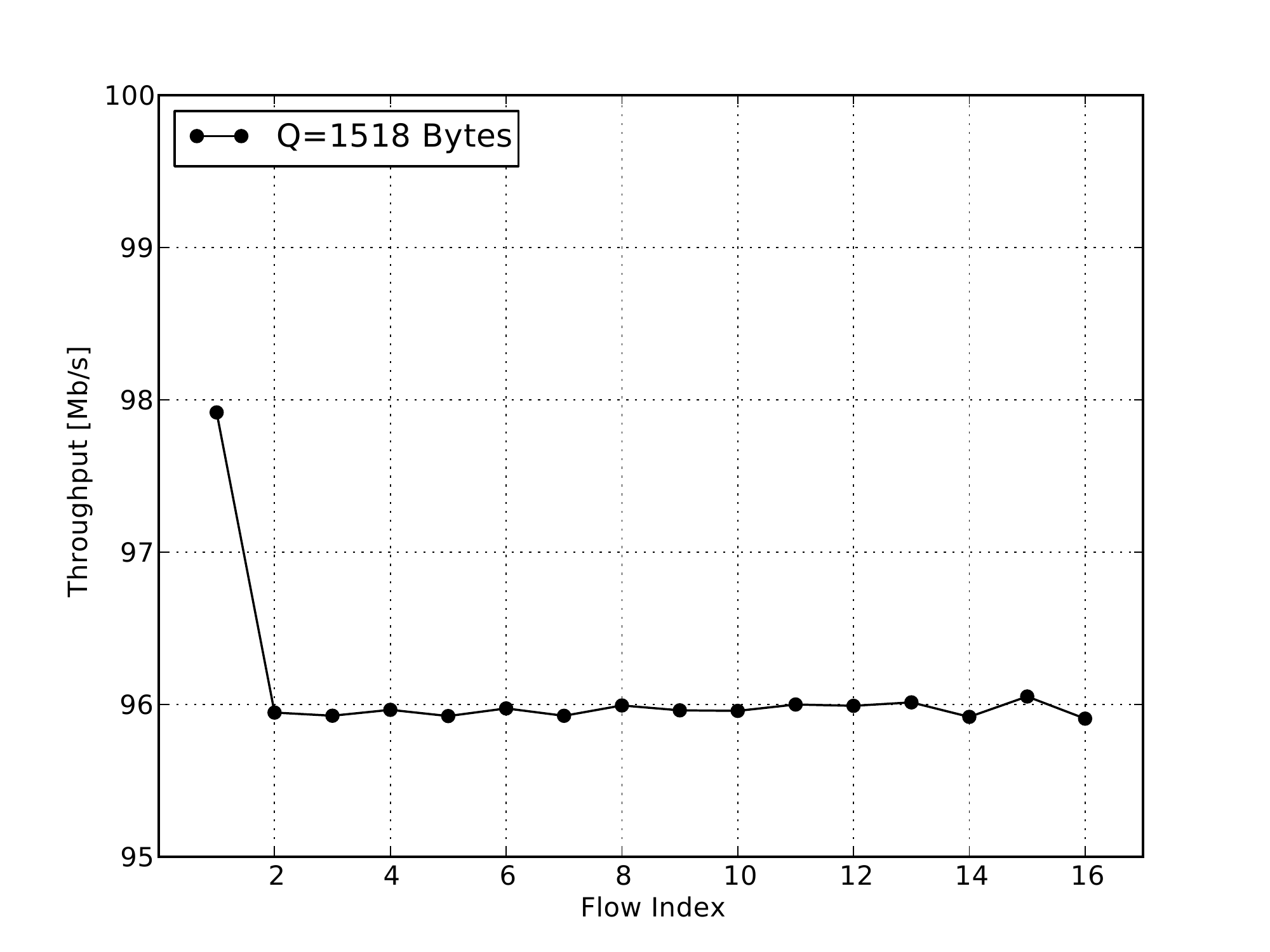}\\
      {\scriptsize (b)}
    \end{center}
  \end{minipage}
  \caption{Throughput for 16 flows with (a) exponential interframe times and
    random frame sizes and (b) exponential interframe times and fixed frame
    sizes.}
  \label{fig:thruput}
\end{figure*}

Fig. \ref{fig:thruput} (a) and (b) show the throughput for 16 flows for two
different sets of conditions for inter-fame times and frame sizes.
%

To demonstrate the performance of the proposed MCDRR scheduling algorithm, we
carried out simulation experiments with a model for a hybrid TDM/WDM link with
tunable transmitters and fixed receivers shown in Fig. \ref{fig:hybrid_link}.

We set the number of wavelengths/channels ($W$), the line rate of each channel,
and the number of tunable transmitters ($M$) to 16, 1 Gb/s, and 2,
respectively. We assume that the scheduling is done at the data link layer with
Ethernet frames and ignore the tuning time of tunable transmitters in
simulation. Each VOQ can hold up to 1000 frames. We measure the throughput of
each flow at a receiver for 10 mins of simulation time.

Fig. \ref{fig:thruput} (a) and (b) show the throughput for 16 flows for two
different sets of conditions for inter-fame times and frame sizes.

In Fig. \ref{fig:thruput} (a), the interframe times are exponentially
distributed with the averages of 16 $\mu$s and 48 $\mu$s for the first flow and
the rest of the flows respectively, while the frame sizes are uniformly
distributed between 64 and 1518 bytes for all the flows. In Fig.
\ref{fig:thruput} (b), the interframe times are exponentially distributed with
the averages of 16 $\mu$s and 32 $\mu$s for the first flow and the rest of the
flows, while the frame sizes are fixed to 1000 bytes for the first flow and 500
bytes for the rest of the flows. For both the cases, the first flow sends frames
at four times the rate of other flows. The combined traffic rates\footnote{The
  inter-fame gap (IFG) of 12 bytes is taken into account.} are 2.409 Gb/s for
Fig. \ref{fig:thruput} (a) and 2.375 Gb/s for Fig. \ref{fig:thruput} (b), which
slightly overload the link. Raj Jain's fairness index \cite{Jain:84} for the
results of Fig. \ref{fig:thruput} (a) and (b) are 0.9999756 and 0.9999998,
respectively.

From the simulation results, we found that the proposed MCDRR scheduling
algorithm provides nearly perfect fairness even with ill-behaved flows for
different sets of conditions for interframe times and frame sizes.

\section{Conclusion}
In this paper we have proposed and investigated the performance of the MCDRR
scheduling algorithm for a multi-channel link with tunable transmitters and
fixed receivers, which is based on the DRR, the well-known single-channel
scheduling algorithm. In extending the DRR to the case of multi-channel
scheduling, we try to efficiently utilize the network resources (i.e., channels
and tunable transmitters) by overlapping rounds, while maintaining its low
complexity (i.e., $O(1)$). The nearly perfect fairness provided by the MCDRR has
been demonstrated through simulation experiments. Establishing mathematical
bounds for the fairness and latency of the MCDRR and comparison with other
multi-channel scheduling algorithms are now under study.

\section{Acknowledgment}
The authors would like to thank the reviewers for their constructive comments.


\end{document}